\documentclass[]{spie}
\usepackage[]{graphicx}
\usepackage{subcaption}

\title{A flux calibration device for the SuperNova 
Integral Field Spectrograph (SNIFS)} 

\author{Simona Lombardo\supit{a}, Greg Aldering\supit{b}, Akos Hoffmann\supit{a},\\
Marek Kowalski\supit{a}, Daniel K\"usters\supit{a}, Klaus Reif\supit{c}, Mickael Rigault\supit{a}
\skiplinehalf
\supit{a}Physikalisches Institut, Nussallee 12, Universit\"at Bonn, Bonn, Germany \\
\supit{b}Lawrence Berkeley National Laboratory, 1 Cyclotron Road, Berkeley, USA \\
\supit{c}Argelander Institute f\"ur Astronomie, Auf dem H\"ugel 71, Universit\"at Bonn, Bonn, Germany \\
}

\authorinfo{lombardo@physik.uni-bonn.de}

\begin{document} 
  \maketitle 
  
\begin{abstract}
Observational cosmology employing optical surveys often require 
precise flux calibration. In this context we present SNIFS Calibration Apparatus (SCALA), a flux calibration system developed for the
SuperNova Integral Field Spectrograph (SNIFS), operating at the
University of Hawaii 2.2$\,$m telescope.
SCALA consists of a hexagonal array of 18 small parabolic mirrors distributed over the face of, and feeding parallel light to, the telescope entrance pupil. The mirrors are illuminated by integrating spheres and a wavelength-tunable (from UV to IR) light source, generating light beams with opening angles of $1^\circ$. 
These nearly parallel beams are flat and flux-calibrated at a subpercent level, enabling us to calibrate our ``$\mathrm{telescope}+ \mathrm{SNIFS}$ system'' at the required precision.  
\end{abstract}



\section{INTRODUCTION}
\label{sec:intro}  

A major goal of modern cosmological surveys is to better constrain the  properties of dark energy. After having proven their importance in detecting the accelerated expansion of the universe\cite{perlmutter_measurements_1999, riess_observational_1998}, type Ia Supernovae (SNe~Ia) remain the strongest demonstrated technique for measuring the dark energy equation of state parameter, $w$. However, current studies are limited by systematic uncertainties, among which the flux calibration is the dominant systematic uncertainty \cite{conley, betoule}.
To reduce this uncertainty, two  issues must be addressed: reaching the 1\% precision photometry required by modern imaging surveys\cite{stubbs1}, and refining the primary standard star network which currently relies heavily on models of white dwarf stars\cite{bohlin}. Accordingly, there is an ongoing effort to develop new techniques and instruments for calibration\cite{dice,regnault}. Here we present a new approach to the problem.

The SNIFS Calibration Apparatus (SCALA) is a flux calibration device developed for the SuperNova Integral Field Spectrograph\cite{lantz} (SNIFS), built by the Nearby Supernova Factory, and mounted on the 2.2$\,$m telescope of the University of Hawaii. SNIFS has a fully filled $6"\times6"$ spectroscopic field-of-view subdivided  by a microlens array into a grid of $15 \times 15$ contiguous square spatial elements (spaxels) forming a 2D grid of spectra. The dual-channel spectrograph simultaneously covers 3200--5200~\AA{} and 5100--10\,000~\AA{} with 2.8 and 3.2~\AA{} resolution, respectively. The purpose of SCALA is to provide an accurate measurement of the instrumental response of the ``$\mathrm{telescope}+ \mathrm{SNIFS}$ system" for each of SNIFS spaxels.
This project is part of the Nearby Supernova Factory (SNfactory) which is producing a large set of flux-calibrated spectrophotometric timeseries of SNe Ia\cite{aldering}.

To achieve the 1\% precision, one needs to control two important factors in the measurement of the brightness of SNe. The first is the atmospheric extinction, which has  already been studied in detail by the SNfactory using a large set of nightly spectroscopic observations of standard stars \cite{buton_atmospheric_2013}. The second is the instrumental response function of SNIFS and the UH 2.2$\,$m telescope, which is currently solved for using the above mentioned observations of standard stars, and solving for the instrument response function. SCALA is an attempt to measure the instrument response function in-situ.

In the next sections we discuss the motivations behind the SCALA concept (Sec. \ref{sec:concept}) and give a detailed description of the device including first test results (Sec. \ref{sec:design}). We  conclude in Sec. \ref{sec:conclusion}.

\section{MOTIVATIONS AND MAIN CONCEPT} 
\label{sec:concept}

The use of SNe~Ia as distance indicators in cosmology is based on the ability to standardize the flux with an uncertainty of 10-15 \% for individual objects. The flux ratio between  nearby and high redshift SNe enables cosmological parameters to be constrained. Photons from distant SNe are redshifted, hence, while the absolute flux calibration cancels in the ratio of brightnesses, one has to establish a reliable (relative) flux calibration as a function of wavelength. 
 SCALA's main purpose is to calibrate the instrumental response of the ``$\mathrm{telescope}+ \mathrm{SNIFS}$''  system to 1\% precision. Such a device will provide an independent  verification of the flux calibration by refining the standard star network. Driven by the requirement for accurate SNe~Ia distance indicators, our current efforts are focused on determining the relative wavelength dependent flux calibration; the absolute flux calibration is left for future work.  

\begin{figure}[]
\centering
\includegraphics[scale=0.45]{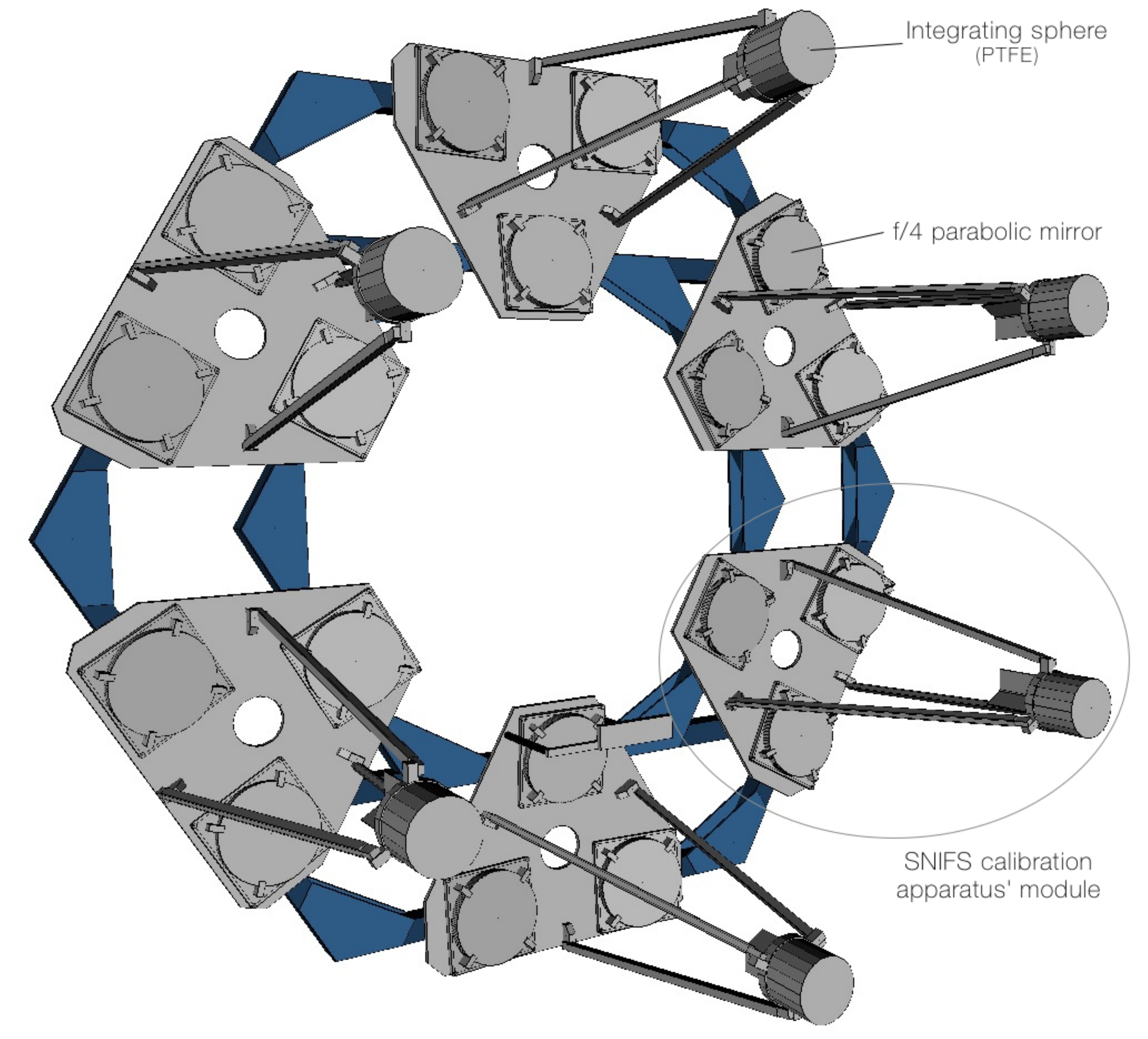}

\caption{Hexagonal arrangment of the six submodules of SCALA. This structure has been mounted in front of the entrance pupil of the telescope.}

\label{Flo:hexagon}
\end{figure}

 We built a device that produces a uniform and homogeneous illumination of the focal plane, or in other words a flat-field, which needs to be flat at a sub-percent level in order to be a subdominant effect. Ideally, a flat-field would be a parallel light beam with the size of the entrance pupil of the telescope (2.2$\,$m). Given the difficulties related with the production of such a light beam we opted for many smaller parallel light beams. SCALA consists of 18 f/4 parabolic mirrors with diameters of 20 cm, distributed over the entrance pupil in a  nearly hexagonal arrangement (see Fig. \ref{Flo:hexagon}). Small integrating spheres, fed  by a wavelength-tunable (from UV to IR) light source, illuminate the mirrors, producing 18 parallel and collimated beams with opening angles of $1^\circ$. The combination of these beams allows us to achieve an illumination of the 2.2$\,$m telescope focal plane that is flat to within 1\%.
 
 A large fraction of the entrance pupil (about 20\%) is sampled uniformly with this configuration. Hence, large scale gradients in the reflectivity of the primary mirror -- as well as small scale variations -- will average out to a large extent. Uniform illumination of the entrance pupil (averaged over roughly 1 m patches) is particularly important for the purpose of SNIFS' calibration. The optics of this spectrograph are such that the PSF at which the individual spectra are imaged on the CCD corresponds to an image of the entrance pupil. Using simulations, we have verified that the chosen sampling reproduces the PSF obtained in the case of entirely uniform illumination of the telescope pupil to a few percent.

\section{DESCRIPTION AND FIRST RESULTS}
\label{sec:design}

In this section, the light source, integrating spheres, projector modules and photodiode systems, and the results of the first tests are discussed.\\

\subsection{Light sources}

The requirement for the light source used in our device is to produce mono-chromatic light of sufficient intensity to allow a calibration measurement with better than 1\% statistical precision over the SNIFS wavelengths range from 320$\,$nm to 1000$\,$nm using a calibrated photodiode (see Sec. \ref{sec:photod}). We have chosen to use a monochromator (Cornerstone 260) for its ease of use, and a 150 W Xenon arc lamp together with a custom-made tungsten halogen lamp to illuminate the monochromator. The former lamp is used from 320$\,$nm to 700$\,$nm and the latter for redder wavelengths. Both lamps are used in order to avoid bright and narrow emission lines produced by the Xenon lamp above 700$\,$nm, obtaining a continuum spectrum from the near UV to the near IR region.
A bandpass of 3 nm was chosen in order to achieve a good balance between resolution and sufficient light level. The light source is entirely computer controlled.

\subsection{Integrating spheres}
\label{issec}

The beam produced by SCALA and imaged in the focal plane of the telescope is an image of the light emitted by the integrating spheres (IS). SCALA's ISs have been designed and produced to achieve a flat field with variations $<1\%$.
 These spherical cavities are coated with white and diffuse reflecting material, usually made of barium sulfate or PTFE (teflon) for use in the visible spectrum. The Barium sulfate coatings are bound with a water soluble glue that makes it difficult to clean and does not adhere well, making it a poor choice. PTFE coatings are made from a powder that produces a sinter inside the sphere. Because these coatings can be expensive and difficult to apply, we instead decided to use a block of PTFE to achieve good diffusion and high reflectivity ($\gg $90\%). In this configuration a light beam entering the IS is scattered through the PTFE block many times before being absorbed or escaping through the exit port. This property produces a homogeneous illumination of the exit port, if it has the appropriate dimension. 

\begin{figure}[h!]
\centering
\includegraphics[scale=0.045]{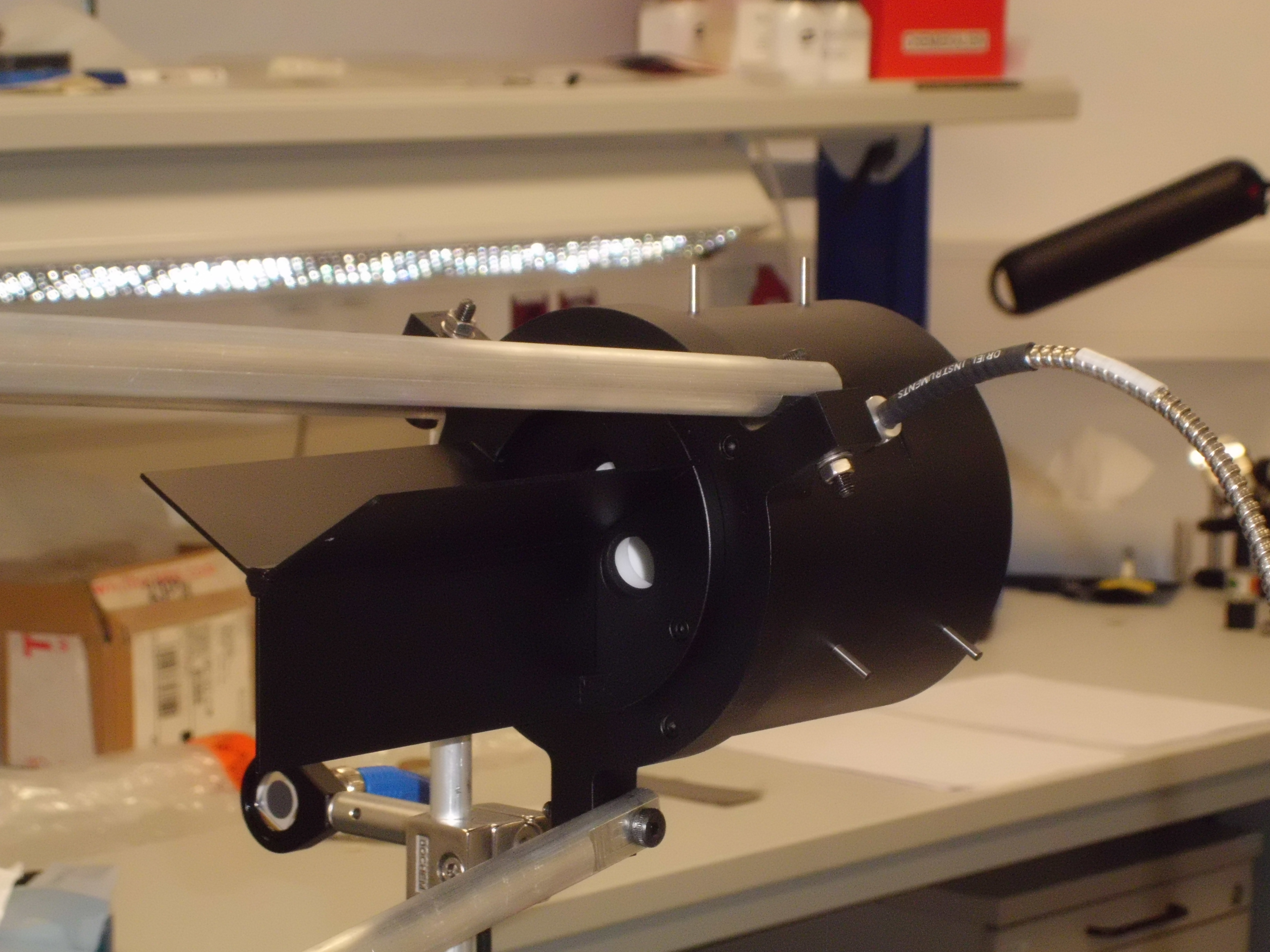}

\caption{One of SCALA's integrating spheres. The cylindrical aluminium housing and the white PTFE inside can be seen through one of the exit ports.}

\label{Flo:ispic}
\end{figure}

The ISs developed for SCALA (Fig. \ref{Flo:ispic}) have three exit ports, illuminating three mirrors in an off-axis configuration. They consist of an outer housing of aluminium with a cylindrical shape. Within this housing we placed the massive PTFE cylinder, constructed by cutting the cylinder in half, machining semispheres on the abutting faces, roughening the semipheres surfaces to improve the diffusion properties, and reconnecting the cylinder.

\begin{figure}[]
\centering
\includegraphics[scale=0.45]{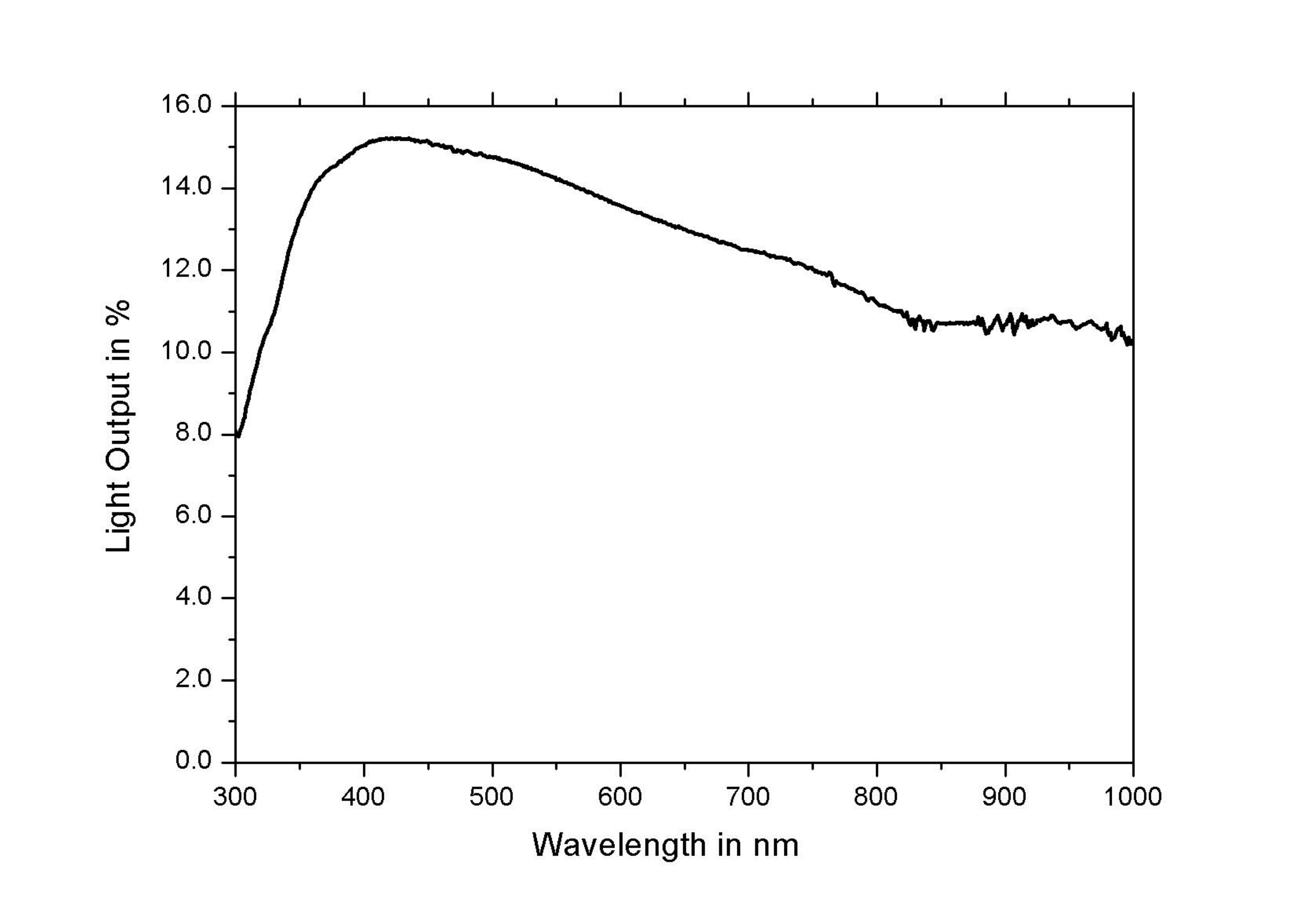}

\caption{Transmissivity of a SCALA prototype integrating sphere. The decrease in the IR is expected because the PTFE becomes partially transparent for large wavelengths.}

\label{Flo:istransm}
\end{figure}

The efficiency of an IS can be estimated according to:\cite{design-IS}
\begin{equation}
\tau=\frac{I_e}{I_i}=\frac{\rho f_e}{1-\rho(1-f_i)}	
\label{eq.eff.IS}
\end{equation} 
where $I_e$ is the emitted flux, $I_i$ is the entering flux, $\rho$ is the reflectivity, $f_e$ is the ratio between the area of the exit port and the area of the sphere, $f_i$ is the ratio between the area of all ports and the area of the sphere. 
In the case of SCALA, we have $f_i=0.02$: the diameter of the spheres is 8$\,$cm and the exit ports are 1.4$\,$cm each. This value of $f_i$ is better than the required 0.05 (which is an upper limit) necessary to have homogeneous illumination of the exit ports.\cite{design-IS}
Fig. \ref{Flo:istransm}  shows the efficiency of SCALA's ISs as a function of wavelength, where we see that $\tau$ is on average 12\%. Thus, according to Eq.\ \ref{eq.eff.IS}
 $\rho=98\%$ .

\subsection{Projector module}

\begin{figure}
\begin{center}
\begin{tabular}{c}
\includegraphics[height=6cm]{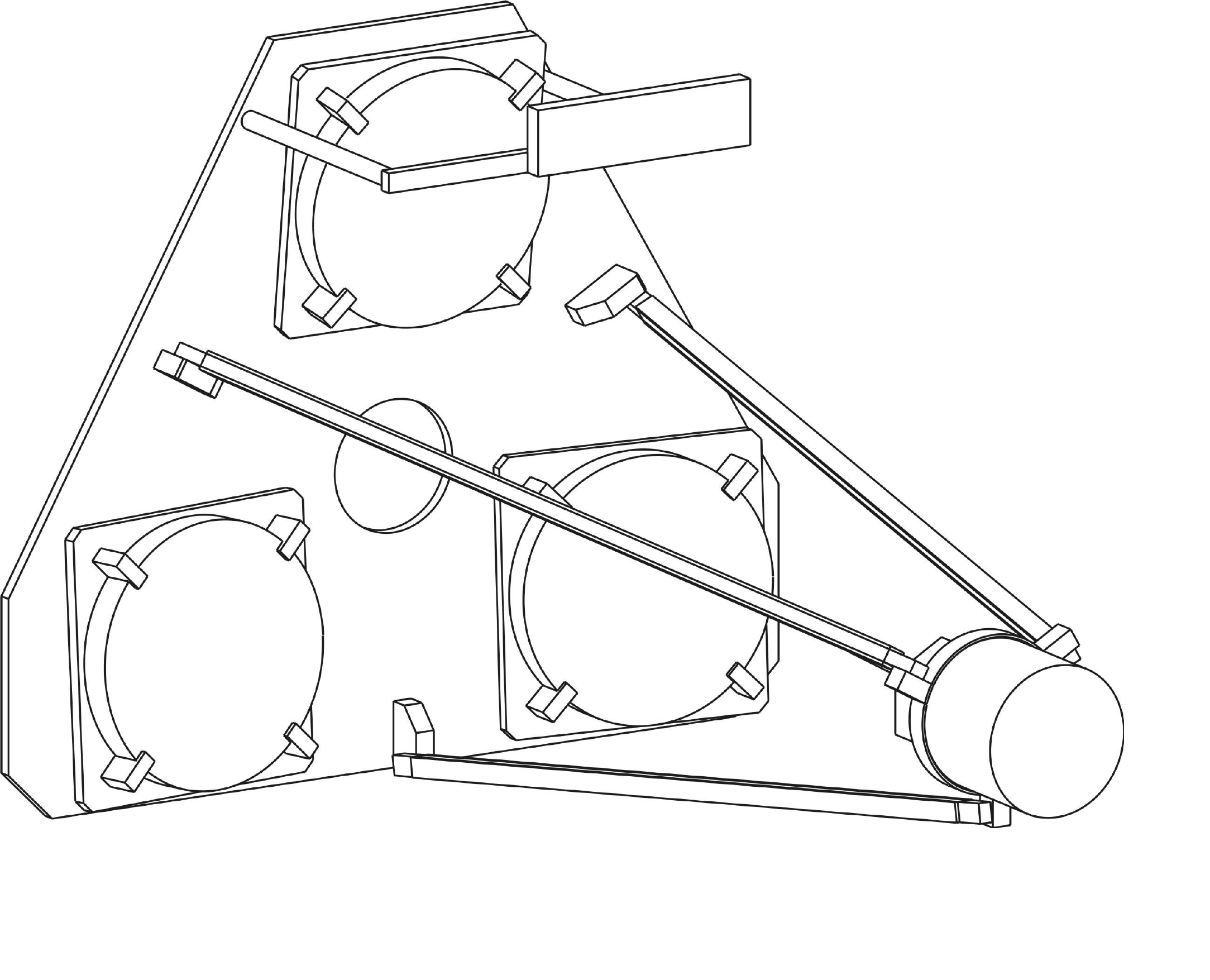}
\includegraphics[height=7cm]{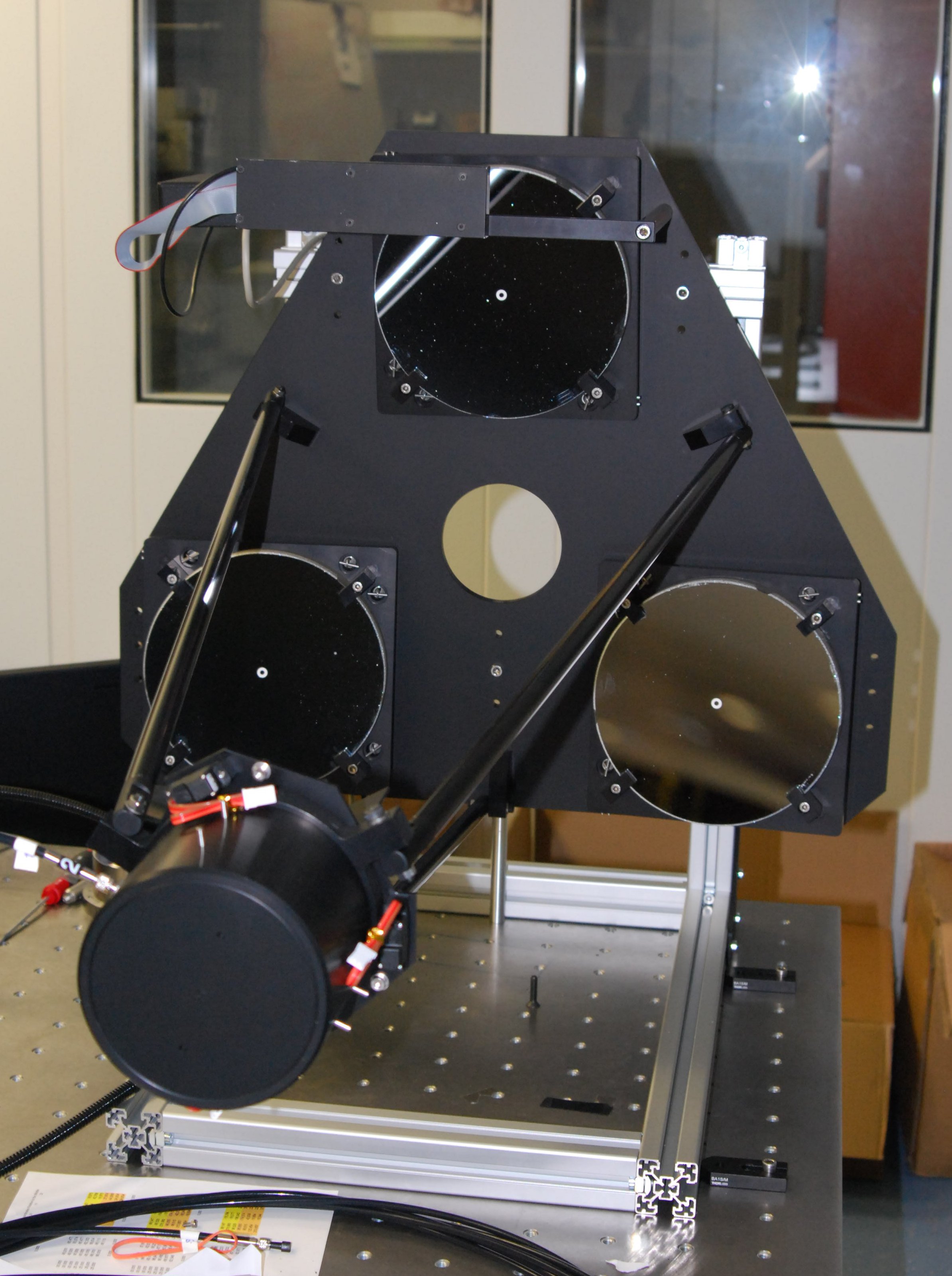}
\end{tabular}
\end{center}
\caption{One of six projector modules for calibration of SNIFS. Left: A projector module with some of the features of	the device: the three mirrors, the integrating sphere (the cylindrical object), and the projector holder. Right: A projector module in the laboratory, the small object mounted in front and towards the upper edge of one of the mirrors is the calibrated photodiode.}
\label{Flo:single}
\end{figure}

SCALA is composed of six projector modules (see Fig. \ref{Flo:single})  arranged in a hexagonal pattern shown in Fig. \ref{Flo:hexagon}. A single fiber bundle, manufactured to fit the slit width and height of the monochromator splits into six bundles that feed the IS. Each of them illuminates three parabolic mirrors, which are mounted on a triangular projector module back structure. Every projector module is therefore composed of an IS and three parabolic mirrors, tilted to produce parallel beams of light with opening angles of $1^\circ$.
The narrowness of the beams reduces the amount of stray light compared to conventional flat fields.  

Most components of SCALA are constructed of aluminium (for a total weight of 150\,kg), and the hexagonal aluminium profiles that hold the projector modules are attached to an iron ladder in the dome of the telescope. Because of this robust construction, SCALA can preserve its position in an environment exposed to wind loads and earthquakes.

\begin{figure}[]
\centering
\includegraphics[scale=0.3]{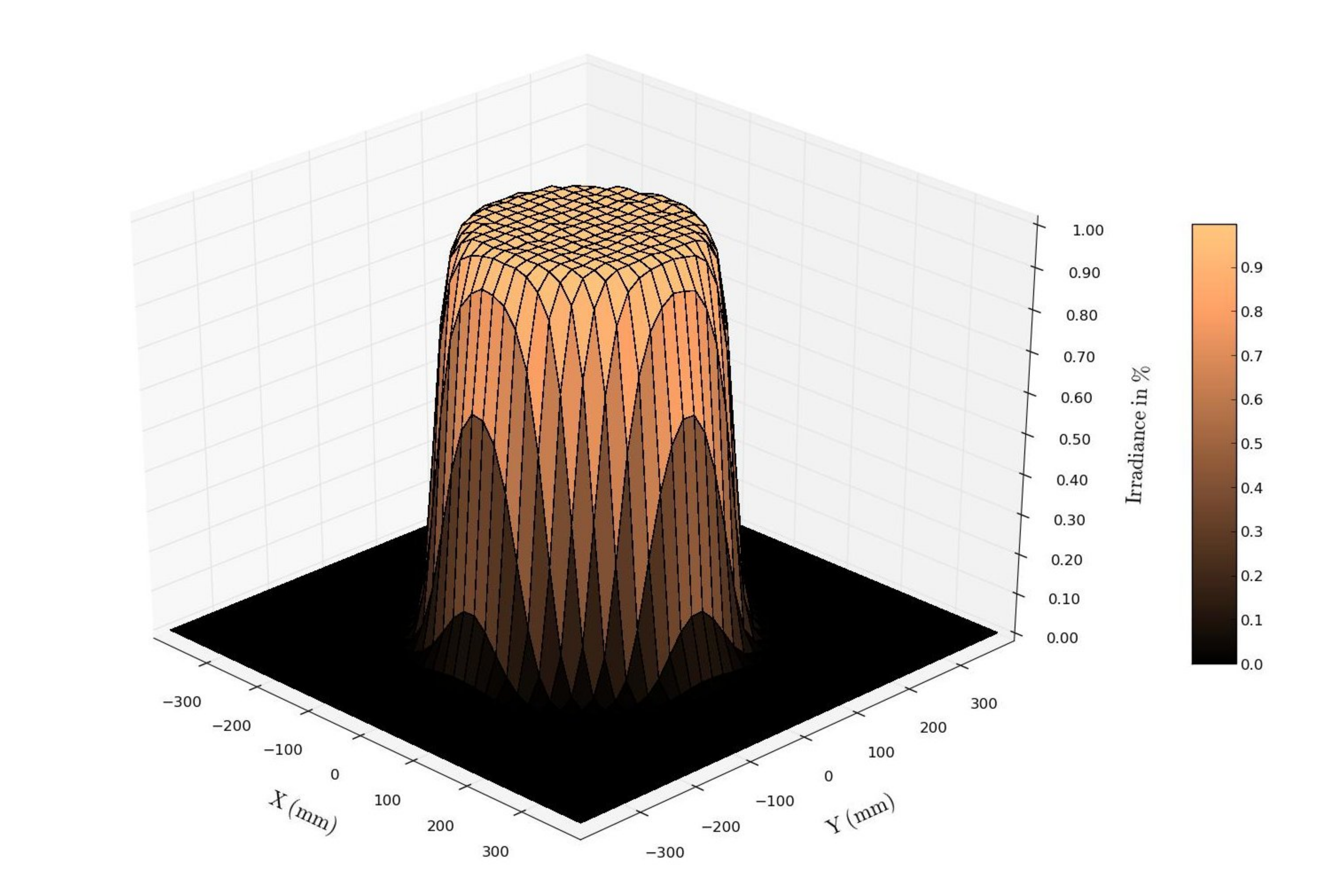}

\caption{Simulation of the illumination of the focal plane of the UH 2.2$\,$m telescope produced by the superposition of the 18 parallel beams. The variations are smaller than 1\% over its full width.}

\label{Flo:beam_fp}
\end{figure}

The off-axis illumination of the 18 mirrors produces a gradient in the single reflected light beam of about 5\%. This gradient is cancelled out because of the symmetrical location of the three mirrors of the projector modules. The overall beam, composed of the overlap of the 18 beams in the focal plane of the telescope, will be even flatter due to the symmetrical illumination of the entrance pupil. The flatness of this design has been verified through simulation (using the photon engineering software FRED\cite{fred}). The simulation result of the final configuration of SCALA is shown in Fig. \ref{Flo:beam_fp}. The beam produced is flat with variations smaller than 1\% over its full width. Considering that a $1^\circ$ beam has a diameter of about 400$\,$mm in the UH 2.2$\,$m telescope focal plane, we can assume that the variations are much smaller over the area of SNIFS ($<1\,$mm, i.e. $6"\times6"$).

\subsection{Calibrated photodiode}
\label{sec:photod}

In order to monitor the amount of light entering the telescope we use two Cooled Large Area Photodiodes (CLAPs\cite{barrelet}) placed in the beams produced by the mirrors. The CLAPs have been developed for (Sn)DICE\cite{regnault} (Direct Illumination Calibration Experiment) and calibrated against a NIST-calibrated photodiode. They have been optimised for our low light levels. The various combinations of fiber bundles, mirrors and integrating spheres result in small spectral variations in the different beams and hence we have to control each of the beams separately.   
The use of 18 CLAPs to calibrate every beam individually was beyond the scope of this project, so we use two CLAPs: the first is used in a stationary configuration to continuously monitor the light level of one of the beams (the reference beam) and a second CLAP is used to calibrate every other beam against the reference beam. We remind the reader that the goal of SCALA is  the relative spectral calibration and our focus is on controlling chromatic effects. Therefore, we do not try to extrapolate the photon flux observed by CLAP to the total photon flux in the beam, which would be required for any attempt of an absolute flux calibration.

\subsection{First tests and results}
\label{sec:results}

\begin{figure}
\begin{center}
\begin{tabular}{c}
\includegraphics[height=8cm]{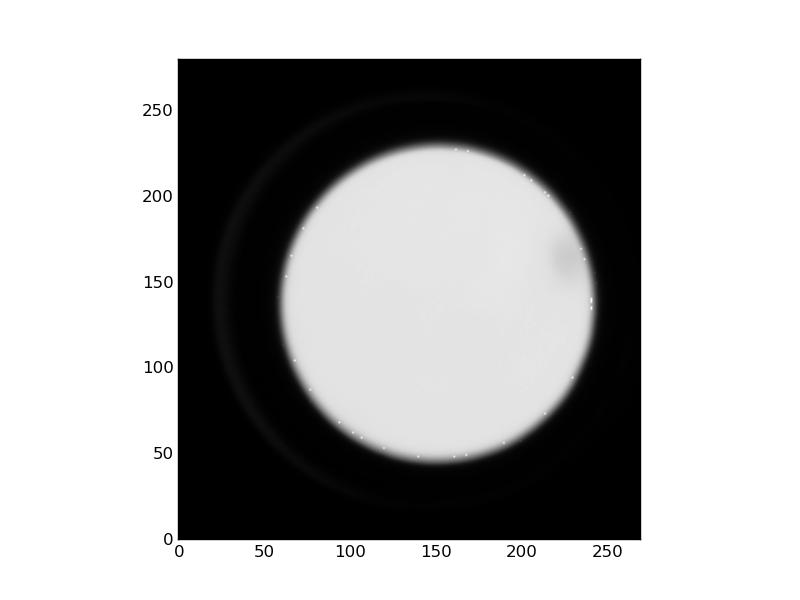}\\
\includegraphics[height=10cm]{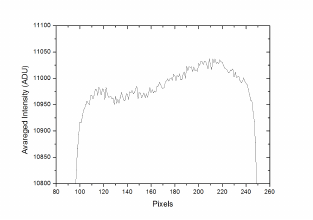}
\end{tabular}
\end{center} 
\caption{Top: the exit port of the IS is aligned with a Atik 383L+ (KAF8300 monochrome) CCD and imaged (the axis are in number of pixels). Bottom: zoom on the averaged intensity of the beam that shows variations of about $0.4\%$ over its full width}
\label{Flo:isfit}
\end{figure}

It is crucial to demonstrate that the overall beam produced in the focal plane of the telescope is flat to within 1\%. Since the beam seen by SNIFS will be an image of the IS exit port, the flatness requirement translates into verifying the uniformity of the light emitted by our integrating sphere.
To ensure this uniformity, we replaced the fiber bundle in the input of the IS with a LED and reimaged the light from the exit port onto a Atik 383L+ (KAF8300 monochrome) CCD.
We aligned an exit port with the CCD, and, after processing the image, we found that the variations are $0.4\%$ over the full width of the beam (see Fig.\ \ref{Flo:isfit}). Considering that SNIFS will see only a very small fraction of this beam, this result shows that the beam fulfils the flatness requirement. 
 \\

\section{CONCLUSION}
\label{sec:conclusion}
We have described SCALA, a device for the in-situ calibration of the SuperNova Integral Field Spectrograph (SNIFS) mounted on the University Hawaii 2.2$\,$m telescope. The purpose of SCALA is  to calibrate the spectra of SNIFS by providing monochromatic flat fields of known chromatic intensity. SCALA consists of 18 mirrors of 20$\,$cm diameter, illuminated by six integrating spheres that are fed by a tunable monochromatic light source.  Located at the telescope pupil, it provides narrow beams of 1 degree  width that illuminate  the focal plane  homogeneously at the sub-percent level with a minimum amount of stray light. The photon flux entering the telescope is monitored using calibrated photodiodes.

 We have shown through simulation and tests that the beam produced in the focal plane of the UH 2.2$\,$m telescope will be flat at the sub-percent level and have verified experimentally that the light emitted by our integrating spheres fulfils the same requirement. All components (lamp, monochromator, mirrors, integrating spheres and fiber bundle) have been characterised in the laboratory. SCALA commissioning is currently ongoing at the UH 2.2$\,$m telescope, where all the properties of the device will again be verified. 

\acknowledgements

We are grateful to Laurent Le Guillou and Nicolas Regnault from LPNHE  for providing the CLAP modules and support. Furthermore we thank  Manuel Danner from the Hamburger Sternwarte for realuminising the mirrors. This work was supported by the German Science Foundation through TRR33 ``The Dark Universe'' as well as  through a grant of the BMBF-Verbundforschung (``Erasmus-F'') and by the Director, Office of Science, Office of High Energy Physics, of the U.S. Department of Energy under Contract No. DE-AC02-05CH11231.

\end{document}